# $\beta$-Ga$_2$O$_3$ Nano-membrane Negative Capacitance Field-effect Transistor with Steep Subthreshold Slope for Wide Bandgap Logic Applications

Mengwei Si, Lingming Yang, Hong Zhou, and Peide D. Ye*

*School of Electrical and Computer Engineering and Birck Nanotechnology Center, Purdue University, West Lafayette, Indiana 47907, United States*

* Address correspondence to: yep@purdue.edu (P.D.Y.)

ABSTRACT

Steep-slope β-Ga$_2$O$_3$ nano-membrane negative capacitance field-effect transistors (NC-FETs) are demonstrated with ferroelectric hafnium zirconium oxide in gate dielectric stack. Subthreshold slope less than 60 mV/dec at room temperature is obtained for both forward and reverse gate voltage sweeps with a minimum value of 34.3 mV/dec at reverse gate voltage sweep and 53.1 mV/dec at forward gate voltage sweep at $V_{DS}$=0.5 V. Enhancement-mode operation with threshold voltage ~0.4 V is achieved by tuning the thickness of β-Ga$_2$O$_3$ membrane. Low hysteresis of less than 0.1 V is obtained. The steep-slope, low hysteresis and enhancement-mode β-Ga$_2$O$_3$ NC-FETs are promising as nFET candidate for future wide bandgap CMOS logic applications.

MAIN TEXT

- **INTRODUCTION**

High temperature solid-state devices and circuits are required for many applications such as aerospace, automotive, nuclear instrumentations and geothermal wells[1,2]. Silicon based complementary metal-oxide-semiconductor (CMOS) technology are not able to operate at such high temperatures, which is limited by its relatively small bandgap of 1.12 eV. CMOS circuits using wide bandgap semiconductors are promising in these high temperature logic applications. Monoclinic β-$Ga_2O_3$ is one of the promising candidates as n-type channel material because of its ultra-wide bandgap of 4.6-4.9 eV and high electron mobility of ~100 $cm^2$/V·s[3-10]. The ultra-wide bandgap can suppress the carrier distribution at the tail of Boltzmann distribution at high temperature. Meanwhile, β-$Ga_2O_3$ also has the advantage to have low-cost native bulk substrates that can be synthesized in large size through melt-grown Czochralski, edge defined film fed growth, and floating zone method[11-16]. To reduce the power consumption in wide bandgap CMOS logic circuits, enhancement-mode operation with threshold voltage ($V_T$) greater than zero and small subthreshold slope (SS) are required, similarly to Si CMOS. Because enhancement-mode operation and small SS reduce both static leakage current and the supply voltage[17]. SS of metal-oxide-semiconductor field-effect transistors (MOSFETs) is limited by the Boltzmann thermal distribution of electrons as 2.3 $k_BT/q$, which is around 60 mV/dec at room temperature. SS would increase much more for conventional MOSFETs operated at high temperatures. Negative capacitance FETs (NC-FETs) have been proposed and attracted much attention to overcome this thermionic limit of SS[18]. In an NC-FET, an insulating ferroelectric material layer is inserted in the gate stack and serves as a negative capacitor so that channel surface potential can be amplified more than the gate voltage, and hence the device can operate with SS less than 60 mV/dec at room temperature. Hafnium zirconium oxide (HZO) is a recently discovered CMOS compatible ferroelectric thin film insulator with the ability to maintain ferroelectricity with ultrathin physical thickness down to less than 2 nm[19-23]. Therefore, the integration of wide bandgap semiconductors and ferroelectric HZO can reduce the thermionic SS degradation at high temperatures and useful to reduce power consumption in high temperature logic applications.

In this paper, we demonstrate β-$Ga_2O_3$ NC-FETs with ferroelectric hafnium zirconium oxide in gate dielectric stack. Subthreshold slope less than 60 mV/dec at room temperature is obtained for both forward and reverse gate voltage ($V_{GS}$) sweeps with a minimum value of 34.3 mV/dec at reverse gate voltage sweep and 53.1 mV/dec at forward gate voltage sweep at $V_{DS}$=0.5 V. Enhancement-mode operation is achieved by tuning the thickness of β-$Ga_2O_3$ with $V_T$ of 0.47 V for forward gate voltage sweep, $V_T$ of 0.38 V for reverse gate voltage sweep and a low hysteresis less than 0.1 V.

- **EXPERIMENTS**

Fig. 1(a) shows the schematic diagram of β-$Ga_2O_3$ NC-FETs, which consists of 86 nm thick β-$Ga_2O_3$ nano-membrane as the channel, 3 nm amorphous aluminum oxide ($Al_2O_3$) layer and 20 nm polycrystalline HZO layer as the gate dielectric, heavily n-doped (n++) silicon substrate as the gate electrode and Ti/Au source/drain as the metal contacts. The silicon substrate was firstly cleaned by RCA standard cleaning and diluted HF dip, to remove organic, metallic contaminants, particles and unintentional oxides, followed deionized water rinse and drying. The substrate was then transferred to an atomic layer deposition (ALD) chamber to deposit 20 nm $Hf_{1-x}Zr_xO_2$ film at 250 °C, using [$(CH_3)_2N$]$_4$Hf (TDMAHf), [$(CH_3)_2N$]$_4$Zr (TDMAZr), and $H_2O$ as the Hf precursor,

Zr precursor, and oxygen precursor, respectively. The $Hf_{1-x}Zr_xO_2$ film with x = 0.5 was achieved by controlling $HfO_2$:$ZrO_2$ cycle ratio to be 1:1. To encapsulate the $Hf_{1-x}Zr_xO_2$ film, 3 nm $Al_2O_3$ was subsequently *in-situ* deposited by using $Al(CH_3)_3$ (TMA) and $H_2O$ as precursors at the same 250 °C, to prevent the degradation of HZO by the reaction with moisture in air. The amorphous $Al_2O_3$ layer is also used for capacitance matching and gate leakage current reduction. The importance of this interfacial $Al_2O_3$ layer on capacitance matching is discussed in detail in Ref. 23. A rapid thermal annealing (RTA) in nitrogen ambient was then performed for 1 minute at 500 °C to enhance the ferroelectricity[23]. Thin β-$Ga_2O_3$ nano-membrane was mechanically exfoliated and transferred to the $Al_2O_3$/HZO/n++ Si substrate from a (-201) β-$Ga_2O_3$ bulk substrate with Sn doping concentration of $2.7 \times 10^{18}$ $cm^{-3}$ (Determined by C-V measurement[7]). Source and drain regions were defined by electron-beam lithography using ZEP520A as e-beam resist. Ar plasma bombardment for 30 s was then applied to generate oxygen vacancies to enhance the surface n-type doping for the reduction of the contact resistance, followed by Ti/Au (30/60 nm) electron-beam evaporation and lift-off processes. Fig. 1(b) shows the false-color scanning electron microscope (SEM) image of the fabricated β-$Ga_2O_3$ NC-FETs with 4 different channel lengths on the same membrane, capturing the β-$Ga_2O_3$ membrane and the Ti/Au electrodes. Fig. 1(c) shows the cross-sectional TEM image of the HZO/$Al_2O_3$ gate stack. All device electrical characterizations were carried out at room temperature with a Keysight B1500 Semiconductor Parameter Analyzer and a Cascade Summit probe station.

- **RESULTS AND DISCUSSION**

Fig. 2(a) shows the polarization versus voltage hysteresis loop (P-V) for TiN/20 nm HZO/TiN capacitor at different annealing temperatures. The P-V shows clear dielectric to ferroelectric transition of HZO after annealing, while the P-V shows weak dependence on annealing temperature above 400 °C. The metal-insulator-metal (MIM) capacitors are used for Landau coefficients (α, β, γ) extraction for ferroelectric HZO layer only. Fig. 2(b) shows the polarization versus voltage (P-V) characteristics for n++ Si/20 nm HZO/3 nm $Al_2O_3$/Ni stack (the same gate stack of the β-$Ga_2O_3$ NC-FETs) annealed at 450 °C at different voltage sweep ranges. The P-V shows clear ferroelectric hysteresis loop. The P-V characteristics of a thin film ferroelectric insulator can be modeled using Landau-Khalatnikov (L-K) equation[18]. The L-K equation for P-V can be expressed as $V_f = 2\alpha t_f P + 4\beta t_f P^3 + 6\gamma t_f P^5 + \rho t_f \frac{dP}{dt}$, where $V_f$ is the voltage across the ferroelectric insulator, P is the polarization charge, $t_f$ is the thickness of the ferroelectric insulator, α/β/γ are the Landau coefficients and ρ is an equivalent damping constant of the ferroelectric insulator. Landau coefficients are extracted directly from the P-V characteristics in Fig. 2(a) on ferroelectric HZO after annealing[23]. The Landau coefficients are extracted by fitting to experimental data using L-K equation (assuming dP/dt=0 for static P-V measurement) to be α=-1.911×$10^8$ m/F, β=4.32×$10^9$ $m^5$/F/$coul^2$, and γ=0 $m^9$/F/$coul^4$, as shown in Fig. 2(c). Note that the P-V calculated from L-K equation shows S-shape where the negative dP/dV is the negative capacitance, as shown in Fig. 2(c). However, this negative dP/dV cannot be observed from experimental P-V (Fig. 2(a) and Fig. 2(b)) because of the negative capacitance state is unstable, which leads to the hysteresis in real experimental P-V measurement. Energy (U) versus charge (Q) is plotted based on experimental Landau coefficients and calculation using L-K equations as in Fig. 2(d). The negative second order derivative ($d^2U/dQ^2$) also indicates the existence of negative capacitance. The energy of ferroelectric capacitor tends to stay at the local minimums of the U-Q so that the negative capacitance (where $d^2U/dQ^2$<0) is unstable. As a result, NC-FETs may exhibit large hysteresis if the unstable negative capacitance effect is too strong. The key design for the β-Ga$_2$O$_3$ NC-FETs in this work is to stabilize the unstable negative capacitance by matching the capacitance of Al$_2$O$_3$ layer (C$_{ox}$) and the depletion capacitance (C$_D$) in β-Ga$_2$O$_3$ layer with the capacitance of the ferroelectric HZO layer (C$_{FE}$). Therefore, low hysteresis and sub-60 mV/dec subthreshold slope at room temperature can be achieved at the same time.

Fig. 3(a) shows the normalized I$_D$-V$_{GS}$ characteristics in log scale of a β-Ga$_2$O$_3$ NC-FET. The back-gate bias is swept from -0.4 V to 2 V in 40 mV per step while the drain voltage (V$_{DS}$) is biased at 0.1 V, 0.5 V and 0.9 V. The whole sweep takes roughly 1 minute. This device has a channel length (L$_{ch}$) of 0.5 μm and channel thickness (T$_{ch}$) of 86 nm, measured by atomic force microscopy (AFM). This particular thickness is chosen to tune the V$_T$ slightly above zero. If the channel is too thick, V$_T$ is negative so that the device becomes depletion-mode while if the channel is too thin, the conducting current is very small[10]. The typical range of physical width of these nano-membrane devices is 0.3~1 μm, determined by scanning electron microscopy (SEM) for the normalization of drain current. The I$_D$-V$_{GS}$ characteristics were measured in bi-direction both forwardly (V$_{GS}$ from low to high) and reversely (V$_{GS}$ from high to low). SS is extracted as a function of I$_D$ for both forward sweep (SS$_{min,For}$) and reverse sweep (SS$_{min,Rev}$) at various V$_{DS}$. Fig. 3(b)-(d) show the SS-I$_D$ characteristics extracted from Fig. 3(a) at V$_{DS}$=0.1 V, 0.5 V and 0.9 V, respectively. The device exhibits SS$_{min,For}$=57.2 mV/dec, SS$_{min,Rev}$=41.0 mV/dec at V$_{DS}$=0.1 V, SS$_{min,For}$=53.1 mV/dec, SS$_{min,Rev}$=34.3 mV/dec at V$_{DS}$=0.5 V, and SS$_{min,For}$=55.0 mV/dec, SS$_{min,Rev}$=34.4 mV/dec at V$_{DS}$=0.9 V. SS below 60 mV/dec at room temperature is demonstrated for both forward and reverse gate voltage sweeps even at relatively high V$_{DS}$. Ga$_2$O$_3$ MOSFETs with 15 nm Al$_2$O$_3$ as gate dielectric exhibit minimum SS=118.8 mV/dec, as shown in supplementary section 1. SS-I$_D$ characteristics at different V$_{DS}$ are similar; slightly better at high V$_{DS}$ due to the larger impact of Schottky barrier at lower V$_{DS}$. Because of the large bandgap of β-Ga$_2$O$_3$, the band-to-band tunneling current at high $V_{DS}$ is suppressed.

Fig. 4(a) shows the $I_D$-$V_{GS}$ characteristics in linear scale of the same β-$Ga_2O_3$ NC-FET as in Fig. 3. $V_T$ is extracted by linear extrapolation at $V_{DS}$=0.1 V for both forward and reverse gate voltage sweeps. $V_T$ in forward gate voltage sweep ($V_{T,For}$) is extracted to be 0.47 V while $V_T$ in reverse gate voltage sweep ($V_{T,Rev}$) is 0.38 V. Hence, enhancement-mode operation with $V_T$ greater than zero for both forward and reverse gate voltage sweeps is demonstrated. A negligible hysteresis is obtained for both on-state (high $V_{GS}$, as shown in Fig. 4(a)) and off-state (low $V_{GS}$, as shown in Fig. 3(a)), except that when $V_{GS}$ is near the threshold voltage region. At the threshold voltage, low hysteresis is achieved to be 90 mV, calculated by using $|V_{T,Rev}-V_{T,For}|$. Note that this hysteresis is negative if we don't take the absolute value because of the gate voltage induced polarization charge inside the ferroelectric HZO. This is in stark contrast to the conventional hysteresis from MOSFETs with interface and oxide traps, where hysteresis is usually positive because of charge trapping in the defect states. The hysteresis of NC-FETs generally comes from two origins. The one is from the unstable negative capacitance state in ferroelectric insulator ($d^2U_{FE}/dQ^2<0$). This hysteresis can be completely removed by well-matched capacitances ($C_{FE}$, $C_{ox}$ and $C_D$) so that $d^2(U_{FE}+U_{ox}+U_D)/dQ^2$ is greater than zero for all Q for total capacitance to remove the unstable negative capacitance state. The stability condition (static non-hysteretic condition) for the 20 nm HZO/3 nm $Al_2O_3$ has been confirmed to be fulfilled in Ref. 23 so that it is not the origin of the hysteresis in the β-$Ga_2O_3$ NC-FETs in this work. The second origin of the hysteresis is a dynamic effect of measurement because of the ferroelectric dumping factor (ρ) in dynamic L-K equations[24], which can explain the hysteresis measured in the β-$Ga_2O_3$ NC-FETs in this work. Fig. 4(b) shows the $I_D$-$V_{DS}$ characteristics of the same β-$Ga_2O_3$ NC-FET with $V_{GS}$ from -0.5 V to 2.5 V in 0.5 V step and $V_{DS}$ swept from 0 V to 2 V. A linear current-voltage relationship at low $V_{DS}$ shows a relatively good contact property at metal/β-Ga$_2$O$_3$ interface. The relative low drain current in this work, compared to Ref. 10, is not clear. The interface situation of β-Ga$_2$O$_3$ on HZO gate stack seems very different from that on SiO$_2$ in Ref. 10. The requirement of thick β-Ga$_2$O$_3$ membrane (60 nm ~ 80 nm) to observe β-Ga$_2$O$_3$ enhancement-mode operation indicates the significant interface traps and surface depletion exhibit. Although the exact mechanism why the interface traps do *not* affect steep-slope observation on β-Ga$_2$O$_3$ NC-FETs is not clear at this moment, we suspect that it is related to the ultra-wide bandgap of β-Ga$_2$O$_3$.

- **CONCLUSION**

Steep-slope β-Ga$_2$O$_3$ negative capacitance field-effect transistors are demonstrated with ferroelectric HZO in gate dielectric stack. SS less than 60 mV/dec at room temperature is obtained for both forward and reverse gate voltage sweeps with a minimum value of 34.3 mV/dec at reverse gate voltage sweep and 53.1 mV/dec at forward gate voltage sweep at $V_{DS}$=0.5 V. Enhancement-mode operation with threshold voltage ~0.4 V is achieved by tuning the thickness of β-Ga$_2$O$_3$ membrane. And a low hysteresis less than 0.1 V is obtained. The steep-slope, low hysteresis and enhancement-mode β-Ga$_2$O$_3$ NC-FETs are promising nFET candidate for future wide bandgap CMOS logic applications.

## ASSOCIATED CONTENT

**Supporting Information**.

The supporting information is available free of charge on the ACS Publication website. Additional details of $Ga_2O_3$ MOSFETs with $Al_2O_3$ only as gate dielectric are in the supporting information.

## AUTHOR INFORMATION

- **Corresponding Author**

*E-mail: yep@purdue.edu

- **Author Contributions**

P.D.Y. conceived the idea and supervised the experiments. M.S. did the ALD of HZO and $Al_2O_3$. H.Z. exfoliated the β-$Ga_2O_3$ membrane. M.S. and L.Y. did the P-V measurement. M.S. performed the device fabrication, electrical characterization and analyzed the experimental data. M.S. and P.D.Y. co-wrote the manuscript and all authors commented on it.

- **Funding Sources**

This material is based upon work supported by the AFOSR under Grant FA9550-12-1-0180 and in part by DTRA under Grant HDTRA1-12-1-0025.

- **Notes**

The authors declare no competing financial interest.

## ACKNOWLEDGEMENTS

The authors would like to thank M. Fan and H. Wang for the technical support and Chun-Jung Su, S. Salahuddin, and K. Ng for valuable discussions.

**Figure captions**

**Figure 1** (a) Schematic view of β-Ga$_2$O$_3$ NC-FETs. The gate stack includes the heavily n-doped Si as gate electrode, 20 nm HZO as ferroelectric insulator, 3 nm Al$_2$O$_3$ as capping layer. 30 nm Ti/60 nm Au are used as source/drain electrodes. 86 nm Sn doped n-type β-Ga$_2$O$_3$ is used as channel. (b) Top-view false-color SEM image of representative β-Ga$_2$O$_3$ NC-FETs on the same membrane with different channel lengths. (c) Cross-sectional view of HZO/Al$_2$O$_3$ gate stack, capturing the polycrystalline HZO and the amorphous Al$_2$O$_3$.

**Figure 2** (a) Polarization versus voltage characteristics for TiN/20 nm HZO/TiN capacitor at different annealing temperature. The P-V shows clear dielectric to ferroelectric transition after annealing. (b) Polarization versus voltage characteristics for 20 nm HZO/3 nm Al$_2$O$_3$ stack annealed at 450 °C at different voltage sweep ranges. (c) Landau coefficients extracted from Fig. 2(a) and the corresponding P-V. (4) Energy versus charge based on experimental Landau coefficients in Fig. 3(c). The negative second order derivative (d$^2$U/dQ$^2$) indicates the existence of negative capacitance.

**Figure 3** (a) I$_D$-V$_{GS}$ characteristics in log scale of a β-Ga$_2$O$_3$ NC-FET. This device has a channel length of 0.5 μm and channel thickness of 86 nm. SS versus I$_D$ characteristics of the same device in Fig. 3(a) at (b) V$_{DS}$=0.1 V, (c) V$_{DS}$=0.5 V, and (d) V$_{DS}$=0.9 V. SS below 60 mV/dec at room temperature is demonstrated for both forward and reverse gate voltage sweeps.

**Figure 4** (a) I$_D$-V$_{GS}$ characteristics in linear scale of the same β-Ga$_2$O$_3$ NC-FET as in Fig. 3. V$_T$ is extracted by linear extrapolation at V$_{DS}$=0.1 V for both forward and reverse sweeps. (a) I$_D$-V$_{DS}$ characteristics of the same β-Ga$_2$O$_3$ NC-FET as in Fig. 3.

**Figure 1.**

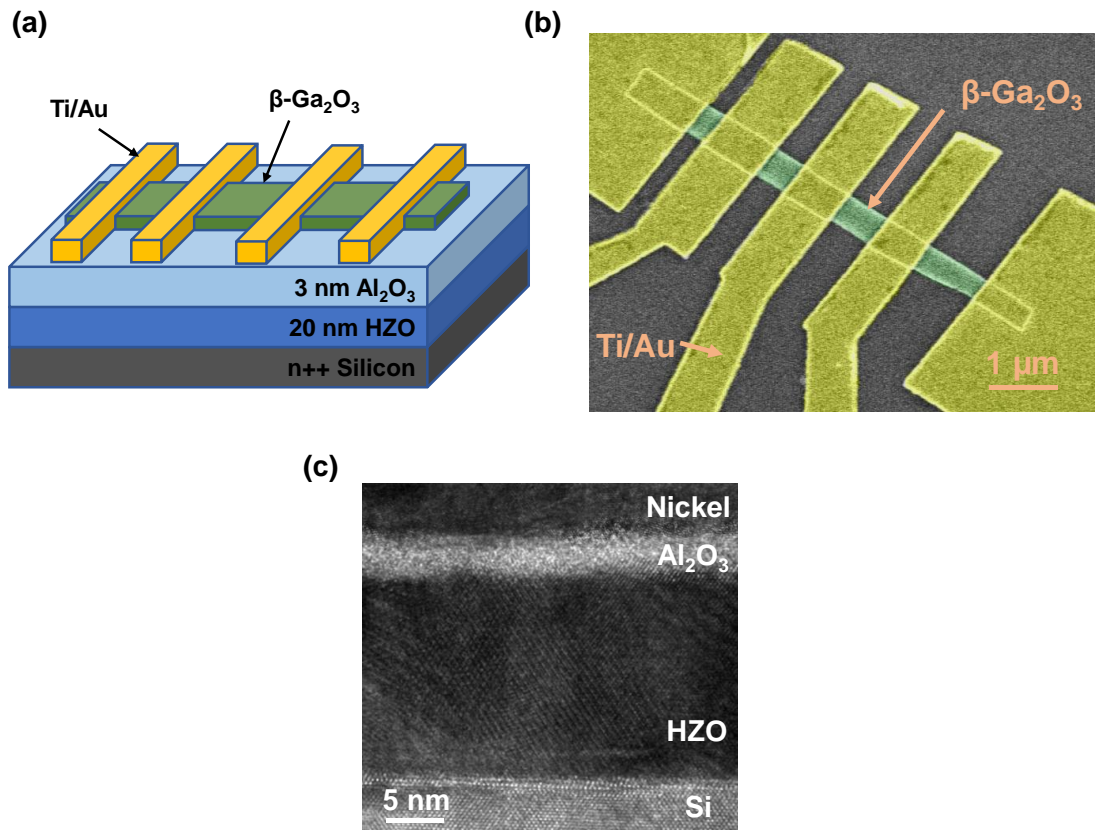

**Figure 2.**

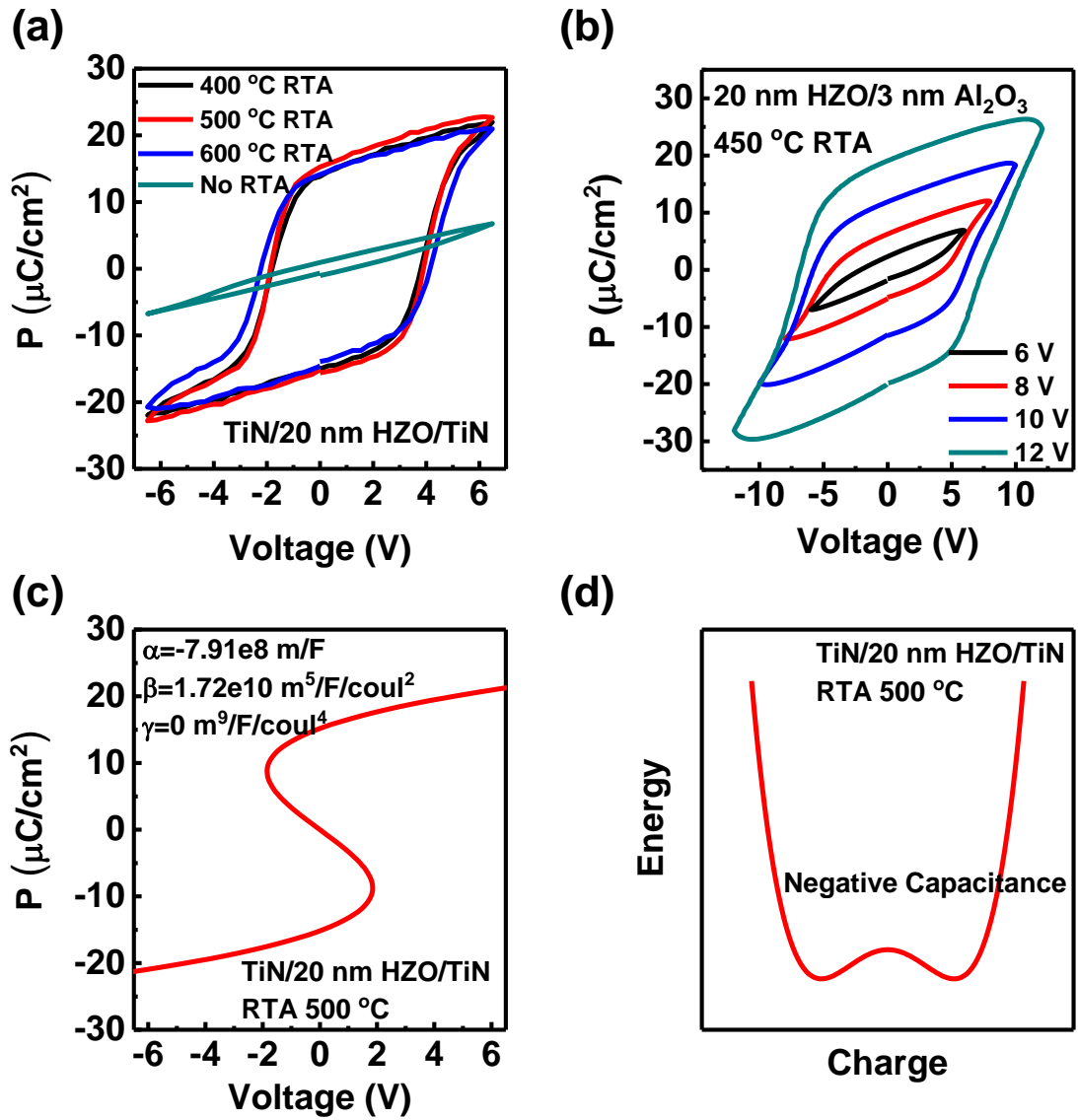

**Figure 3.**

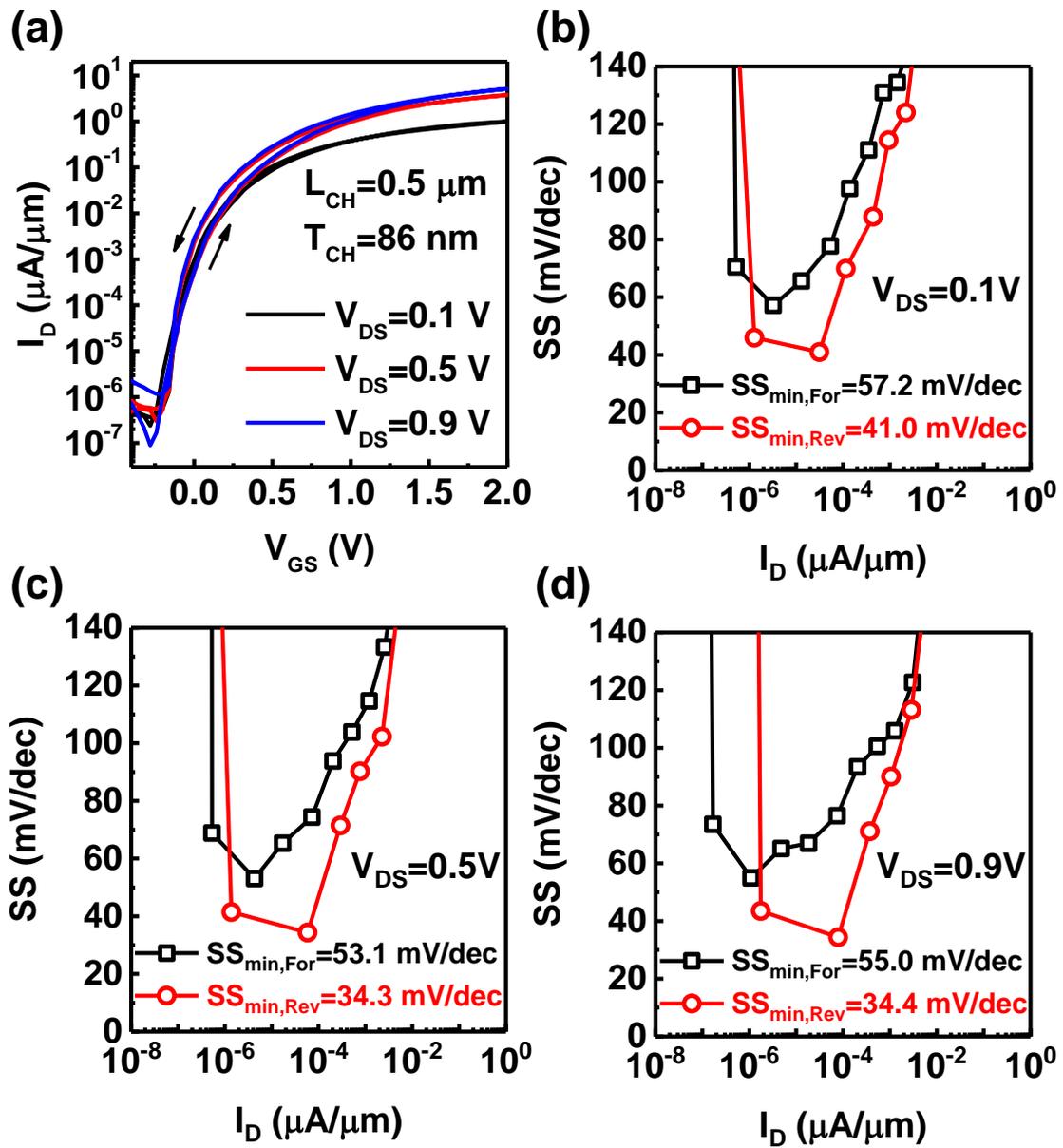

**Figure 4.**

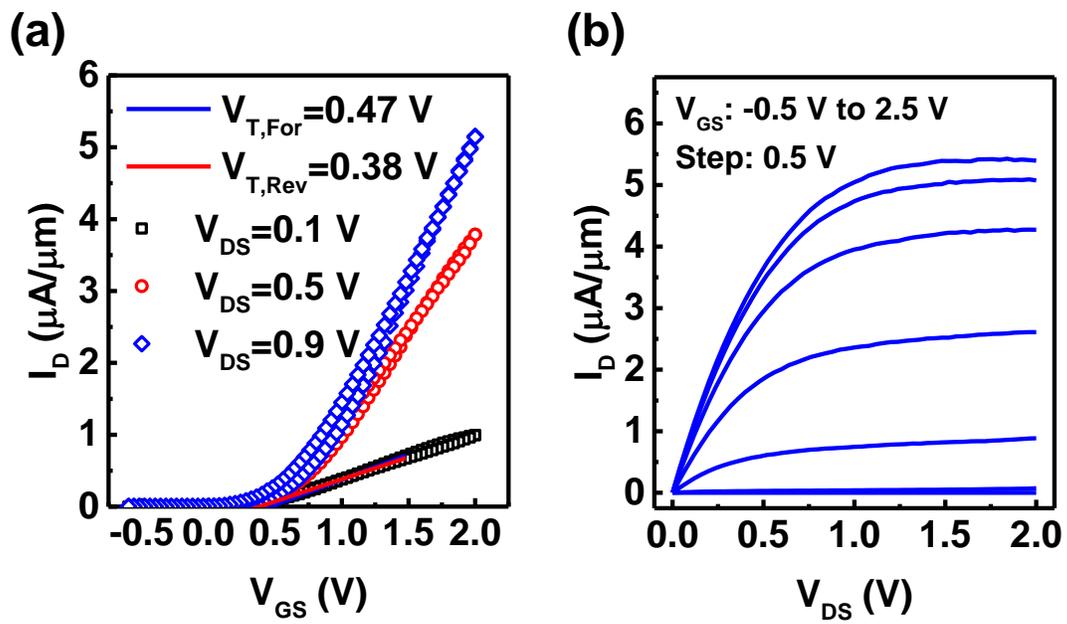

Supplementary Information for:

# $\beta$-Ga$_2$O$_3$ Nano-membrane Negative Capacitance Field-effect Transistor with Steep Subthreshold Slope for Wide Bandgap Logic Applications

Mengwei Si, Lingming Yang, Hong Zhou, and Peide D. Ye*

*School of Electrical and Computer Engineering and Birck Nanotechnology Center, Purdue University, West Lafayette, Indiana 47907, United States*

* Address correspondence to: yep@purdue.edu (P.D.Y.)

## 1. Ga₂O₃ MOSFET with 15 nm Al₂O₃ as gate dielectric

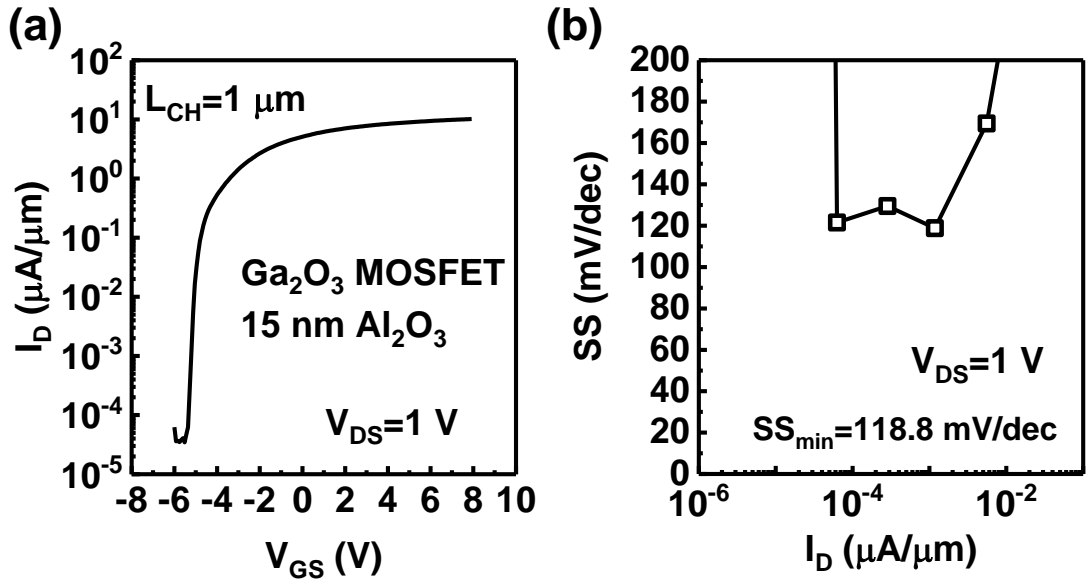

**Figure S1**. (a) $I_D$-$V_{GS}$ characteristics in log scale of a β-Ga₂O₃ MOSFET. This device has a channel length of 1 μm. (b) SS versus $I_D$ characteristics of the same device in Fig. S1(a) at $V_{DS}$=1 V.

Ga₂O₃ MOSFETs with 15 nm Al₂O₃ only as gate dielectric are fabricated to compare with Ga₂O₃ NC-FETs. Fig. S1(a) shows the $I_D$-$V_{GS}$ characteristics of a representative Ga₂O₃ MOSFETs measured at $V_{DS}$=1 V. Fig. S2(b) shows the SS versus $I_D$ characteristics of the same devices with minimum SS=118.8 mV/dec.